\def\year{2015}
\documentclass[letterpaper]{article}
\usepackage{aaai}
\usepackage{times}
\usepackage{helvet}
\usepackage{courier}
\usepackage{graphicx,url}
\usepackage{dcolumn}
\usepackage{subfigure}
\usepackage{multirow}
\usepackage{multicol}
\usepackage{float}
\usepackage{color}
\usepackage{stfloats}
\usepackage{epsfig}
\usepackage{epsf,psfrag}
\usepackage{rotating}
\usepackage{fancyhdr}
\usepackage{amssymb}
\usepackage{amsmath}
\usepackage{balance}  
\usepackage{graphics} 
\usepackage{times}    
\usepackage{url}      
\usepackage{makecell}
\usepackage[
  bookmarks=false,
  pdfpagelabels=false,
  hyperfootnotes=false,
  hyperindex=false,
  pageanchor=false,
]{hyperref}
\begin{document}

\title{Understanding Musical Diversity via Online Social Media \\ \scalebox{.7}{[Please cite the ICWSM'15 version of this paper]}}


\author{Minsu Park\thanks{The majority of this work was done while Minsu Park was a research intern at Qatar Computing Research Institute.}$^1$, Ingmar Weber$^2$, Mor Naaman$^1$, Sarah Vieweg$^2$\\ $^1$Jacobs Institute, Cornell Tech\\ $^2$Qatar Computing Research Institute (QCRI)\\ \{minsu, mor\}@jacobs.cornell.edu\\ \{iweber, svieweg\}@qf.org.qa}

\maketitle

\begin{abstract}
\begin{quote}
Musicologists and sociologists have long been interested in patterns of music consumption and their relation to socioeconomic status. In particular, the Omnivore Thesis examines the relationship between these variables and the diversity of music a person consumes. Using data from social media users of Last.fm and Twitter, we design and evaluate a measure that reasonably captures diversity of musical tastes. We use that measure to explore associations between musical diversity and variables that capture socioeconomic status, demographics, and personal traits such as openness and degree of interest in music (into-ness). Our musical diversity measure can provide a useful means for studies of musical preferences and consumption. Also, our study of the Omnivore Thesis provides insights that extend previous survey and interview-based studies.
\end{quote}
\end{abstract}


\section{Introduction}

\noindent The cultural and social significance of music is universal; music is found in every known human culture, and plays a role in rituals, wars, ceremonies, work, and everyday life~\cite{wallin2001origins}. Tia DeNora~\cite{denora2000music} noted that ``Music is not merely a meaningful or communicative medium. It does much more than convey signification through non-verbal means. At the level of daily life, music has power. It is implicated in every dimension of social agency.'' As social media become more ingrained in our lives, it follows that connections between social media use, and habits and norms regarding music consumption, will occur. In this paper, we present an empirical analysis of social media data as they relate to and reveal details of users' musical tastes.

A person's musical consumption can reveal a lot about their personality, preferences, and sense of self. One can have limited tastes; they may listen to a single genre like pop or rap, and not diverge into other genres. On the other hand, another individual may be eclectic in their musical choices and have a playlist filled with jazz, hip-hop, indie rock, classical, and so forth. We often think of such differences as a matter of individual choice and expression; however, to a great degree, it is hypothesized and tested that the diversity of musical tastes can be explained by external factors. For example, previous research has identified a relationship between musical tastes and social factors, and produced the \textit{cultural omnivore thesis}. This thesis describes ``a shift in the orientation of high-status individuals toward an inclusive range of musical preferences that traverses the traditional boundaries between \textit{highbrow}, \textit{middlebrow}, and \textit{lowbrow} genres~\cite{peterson1992understanding,peterson1997rise,peterson2005problems}.'' However, symbolic boundaries between musical genres have been eroding~\cite{goldberg2011mapping} in recent years, which provides an opportunity to rethink the high-to-lowbrow cultural categories in relation to musical diversity. This can lead to a better understanding of the impact of social conditioning on diverse musical tastes, and by proxy, a better understanding of the connection between socioeconomic status, demographics, and the diversity of musical preferences.

To date, the social computing community has examined online listening activity as source of information and recommendations for music~\cite{bu2010music,zheleva2010statistical,farrahi2014impact,turnbull2014using}. However, computational tools and online outlets such as social media can make further contributions toward understanding human behavior related to musical consumption and help to elaborate user-centric music retrieval systems by analyzing personal characteristics. We focus on exploring a new means of measuring the diversity of individual musical tastes by using data collected from social media, and examine the relationship between musical diversity and various individual factors including socioeconomic and demographic information, as well as social and individual information that can be collected from social media.

Through a multi-platform analysis of a dataset of U.S.\ Last.fm\footnote{Last.fm is a music recommendation service. The site builds a detailed profile of each user's musical consumption by recording details of the tracks the user listens to, either from Internet radio stations, or the user's computer or many portable music devices. It also offers some social networking features such as recommending and playing artists to Last.fm friends~\cite{wiki2015lastfm}.} users and their corresponding Twitter accounts, we examine music consumption together with demographics (e.g., age and gender) and other descriptive variables for a community music fans who have an online presence. Using Twitter-derived information for these users, we inferred their socioeconomic information (e.g., income, education level, and area of their residence) as well as other social and personal variables (e.g., how diverse their friends and interests are, and how `open' and `into music' they are). We then defined a measure for musical diversity by applying the notion of shared understanding as socially perceived distances between genres. We suggest that designing a diversity measure can provide a useful means for studies in recommendation systems. Moving from designing a measure to analysis of associations between diversity and individual factors, we suggest this type of analysis can provide meaningful insights that are complementary to those provided by previous survey and interview-based studies regarding the musical omnivore thesis. Our main contributions therefore are as follows:

\begin{itemize}
\item We propose and validate a novel diversity measure that borrows the concept of Rao-Stirling diversity for music consumption. While recent studies~\cite{hurley2011novelty,farrahi2014impact} define diversity (as it relates to music consumption) as the total number of unique genres associated with all artists listened to, we go into more detail, and define diversity as a multidimensional property that has three main attributes: \textit{variety} (the number of unique genres one listened to), \textit{balance} (the listening frequency distribution across these genres), and \textit{disparity} (the degree of distance between musical categories).

\item We investigate the relation between musical diversity and various other variables including socioeconomic factors. In particular, we find that followers of high-profile news media are more likely to have diverse musical tastes. We also consistently find a weak, but robust trend for people who are more `into' music to have less diverse tastes. Along with these findings, our results also show that demographic factors such as age and gender are associated with musical diversity rather than conventional socioeconomic status such as income and education level.
\end{itemize}

We begin by reviewing the primary key research around the diversity of musical tastes, and then identify possible challenges for developing better measures of diversity.


\section{Related Literature}

\noindent Disciplines such as sociology and social computing addressed the notion of \textit{cultural omnivorism} and the importance of understanding the musical diversity. Given the wealth of related work on these topics, our review focuses on what could be tested by complementing the limitations of previous studies through social media data and how we can design a meaningful measure for the diversity of musical tastes.

\subsection{Changing Status of the Omnivore Thesis}

\noindent Since the publication of Bourdieu's seminal work \textit{Distinction}~\cite{bourdieu1984distinction}, in which he explains the notion of cultural capital and exhibits how access to education, knowledge of the arts, and familiarity with other highly regarded aspects of western culture lead to a `highbrow' status, copious research has investigated the relationship between socioeconomic position and musical tastes~\cite{coulangeon2007distinction}. The majority of the current studies on the \textit{omnivore thesis} in relation to musical tastes, proposed by Richard Peterson~\cite{peterson1992understanding} show that people with a higher socioeconomic status have broader (omnivorous) musical tastes than those with a lower socioeconomic status who have limited (univorous) musical preferences in lowbrow music. There are generally two definitions of omnivorousness, referred to as the \textit{volume} and the \textit{compositional} definitions~\cite{warde2007understanding}. The first refers to higher socioeconomic status people favoring more musical genres than those of lower socioeconomic status. The second refers to the situation that people with higher socioeconomic status tend to have more eclectic tastes across the spectrum of high-to-lowbrow music than people with lower socioeconomic status.

More recently, however, Peterson~\cite{peterson2005problems} conducted comparative research and noted that ``despite the attention paid to the concept by numerous scholars, the subtypes of omnivorousness suggested by them were diverse and fall into no recurrent patterns due to changes in the socio-cultural world.'' Indeed, though there is a little disagreement that the contemporary era has witnessed shifts in the ways cultural preferences and practices are mapped onto social locations, the extent to which this implies changes in the functioning of cultural capital remains unclear~\cite{rimmer2012beyond}. In addition, Peterson~\cite{peterson2005problems} raised a question regarding the traditional measurement of omnivorousness, and recent qualitative studies identified a number of limitations in conventional survey-based studies~\cite{warde2007understanding,rimmer2012beyond}: First, the simple or compositional volume of genres preferred by an individual is insufficient to show the full picture of one's form of engagement and social status since different conceptual frameworks may provide different understandings. Second, there is a tendency to discriminate genres within preferred genres (i.e., even though one answers `rock' as a preferred genre, it does not mean that one likes \textit{all} kinds of rock; therefore, it is possible that someone who likes a Heavy Metal, a subgenre of rock, says ``I like rock,'' and someone who likes the same subgenre says ``I don't like rock''). This inability to discriminate genres, or lack of knowledge regarding how to best express what genres one prefers, can create confusion~\cite{rentfrow2003re}. This gap may bring inconsistency in the preference scoring across survey participants. Finally, the high-to-lowbrow scheme should be reconsidered in contemporary social contexts as Peterson (2005) argues that there is no consensus. In addition, a lot of research has used inconsistent levels of genres, e.g., a questionnaire of preferences for opera, jazz, rock, and heavy metal may be used in these types of surveys, even though heavy metal is often considered a subgenre of rock.

We believe online social media data can help rectify some of these limitations and provide a unique and useful perspective on the musical omnivore thesis: data collected from social media sites can provide a unique capacity to (i) reduce the inconsistency of preference scoring (which may differ across people due to their inability to discriminate) by systematically classifying the genres consumed by users, (ii) explore a different level of relationship between social status and musical tastes by accessing the subgenres of choice among users, which are more fine-grained than higher-level genres, and (iii) analyze data on a consistent level of genre-hierarchy. Further, social media data can provide users with open-ended spaces~\cite{lewis2012social} in which to list their favorite music, concert attendance, and direct/indirect musical information sources, which offers an unprecedented opportunity to examine how tastes are associated with various individual factors. Up to now, the majority of research on musical tastes has relied on closed-ended surveys typically measuring preferences in terms of genres, and our aim is to contribute a new way to look at the relationship between musical preference and various social and individual factors.

\subsection{Technology and Music Listening Practice}

\noindent Exploring musical diversity is an interesting challenge in social computing, as well as music information retrieval (MIR); it also has many applications in real-life scenarios. In MIR, some researchers have explored to achieve the optimal balance between the two objectives on recommendation, similarity and diversity, because it has been recognized that being accurate with similarity metric alone is not enough to judge the effectiveness of a recommendation system~\cite{mcnee2006being,chen2013personality}. In addition, recent studies~\cite{chen2013personality,farrahi2014impact} suggest that one's personality might have a role in the formation and maintenance of music preferences, and diversity of musical tastes could serve as a proxy of the level of openness of one's personality. These studies show that looking at musical diversity as an indicator of openness can have an impact on the performance of a collaborative filtering recommender system. In social computing, diversity has been considered in studying phenomena such as peer influence and music consuming mechanism. Some of this research confirms that informational influence is the key underlying mechanism of music listening practices~\cite{yang2014effect} and systematic recommendations affect users' choices of music tracks and listening behaviors~\cite{buldu2007complex}.

\subsection{Research Questions}

We believe associations between musical categories (e.g., genre-to-genre and subgenre-to-subgenre) can be reasonably derived from the perception of crowds by analyzing their musical consumption, and these distances may help design better measures of musical diversity. The existing measures, \textit{volume} or \textit{entropy}, are different from diversity, and thus cannot accurately capture its essence. Volume, which is defined as the number of musical categories one listens to, does not consider whether a person listens with balance. A 99\%-1\% split between two genres would be treated the same as a 50\%-50\% split. Entropy, on the other hand, takes the distribution into account, so a more skewed distribution would be considered less balanced. However, entropy does not look at the similarities of the musical categories and implicitly assumes all categories to be equidistant to each other (e.g., listening to three different styles of metal music would be the same as listening to classical music, death metal, and salsa). People, however, do consider certain types of music as similar or dissimilar~\cite{morchen2005databionic}. To define and to quantify this notion of similarity we use \emph{co-consumption} behavior. For example, if both rap and hip-hop are consumed by many people we assume that these two genres are similar. Having musical consumption data for a large user set can reveal the distance between musical categories.

The challenges and opportunities in studying musical diversity lead us to introduce two research questions that guide the remainder of this paper:

\begin{description}
\item[RQ1] \textit{Can a novel diversity measure using variety, balance, and distance between musical categories capture the diversity of musical tastes better than existing methods?}

\item[RQ2] \textit{What variables are associated with diversity in music consumption? Is socioeconomic status a factor or are other factors also associated?}
\end{description}


\section{Method}

\noindent The literature referenced in the previous section points to three major dimensions of explanatory variables: socioeconomic status, demographic information, and `openness' (degree of appreciation for novelty and variety of experience). With these dimensions and the additional dimension of `into-ness' (degree of self-disclosed interest in music) as a guide, we identified 15 variables. We inferred socioeconomic status including income, education level, ethnic diversity of area of residence, and urbanness of area of residence by using geocoded tweets. Into-ness (i.e., degree of music-related topics of interest in Twitter) and openness including number of friends, timezone diversity of friends, and interest diversity was inferred by using tweets, profile descriptions, and friendship information in Twitter. We directly downloaded demographic information (e.g., gender and age) and other types of into-ness (e.g., number of event attendance in the past, number of loved tracks, period after registration, and number of friends in Last.fm) through the Last.fm API.

\subsection{Initial Data Collection}

\noindent To identify and obtain a sample of Last.fm users in the U.S.\ who share gender, age, and Twitter user names in their Last.fm profiles, we used the Google Custom Search API and the Bing Search API. We created a custom query containing parameters that returned only Last.fm user pages which contained this particular information. To augment the sample size, we collected U.S.\ Twitter users who share their Last.fm accounts in their Twitter profiles by using the `Search Bio' feature in Followerwonk\footnote{\url{https://followerwonk.com}}. This allowed us to obtain 23,294 unique users. Then, we collected all publicly available tweets from that user population. During this process 4,392 unique users were screened out since some of them did not allow public access to their tweets or had removed their accounts in the meantime. This left us with 18,902 unique users. To infer socioeconomic status by using geocodes in tweets, we limited our remaining sample to those users who posted at least ten tweets with geocodes, which resulted in 3,548 users. Along with Twitter data, we collected Last.fm data including `Top artists' list (i.e., the 50 musicians a user listened to the most; listening frequency for each artist is included) as well as demographic and some into-ness information directly through the Last.fm API.

\subsection{Socioeconomic Status}

\noindent We used home location derived from Twitter as an index to approximate socioeconomic data, and news interests, expressed via Twitter's following network, as another proxy for socioeconomic status. 

A user's home location can be a marker of their socioeconomic status. In particular, the socioeconomic status of social media users can be estimated by extracting the users' hometown ZIP codes and matching that to the median ZIP code household income according to the Census Bureau~\cite{lewis2012social}. In addition, using the inferred home location we can check whether a user lives in an urban or rural area~\cite{hecht2014tale}. 

To obtain the home location for a user, we followed a procedure that involved three different methods of identifying a user's possible home ZIP code. We first reverse-geocoded all the latitude and longitude tags for the user into the ZIP codes, using the Nominatim API\footnote{\url{http://www.nominatim.org}}. We also extracted Federal Information Processing Standard (FIPS) codes, which represent specific regions in counties, using the Coordinates to Political Areas API in Data Science Toolkit\footnote{\url{http://www.datasciencetoolkit.org}}. Using the ZIP code data for the user, we inferred a probable home location of a user when we found an intersection between the sets of potential ZIP codes for the user computed by three different methods, the \textit{plurality} and \textit{n-days} methods summarized in \cite{hecht2014tale} and the \textit{plurality with time limitation} described in \cite{castelli2009extracting}.

The plurality approach~\cite{hecht2014tale} assumes that the single region in which a user was the most active is the user's home location. Using this approach, we find the user's mode ZIP code(s) from which tweets were most frequently posted. The \textit{plurality with time limitation} method is based on the finding in~\cite{castelli2009extracting}, that people are most likely home between 10pm -- 6am. Using these parameters, we identify the user's mode ZIP code(s) from which tweets were most frequently posted during that time period. Since the plurality approaches may not be appropriate for users who travel frequently, the final method we used identified the ZIP code(s) in which a user posted over a period of at least 10 days, considering them `local' to that area if they did.

We selected a single home ZIP code (and FIPS code) for each user by intersecting the ZIP code sets resulting from the three methods mentioned above. The final set of users with non-empty intersection had 1,306 users (there were 3,451, 3,258, and 1,822 users with non-empty sets for each of plurality, plurality with time constraint, and n-days methods respectively). All other users for which we could not robustly estimate a location were removed from the data.

Finally, to extract socioeconomic data, we used each ZIP code to query the 2010 US Census data to determine income, education level, and ethnic diversity in the area. We matched each FIPS code to NCHS data for urban--city classification of the area which places every U.S.\ county on a discrete scale from 1 (a large central metro area) to 6 (a sparse rural area). For each user we thus have values for median household income, percentage of bachelor degrees, proportion of white people\footnote{We tested relation between white ratio and `racial and ethic diversity' by using the Ethnic/Racial Diversity Index which defines racial and ethnic diversity as $1 - \sum_{r \in G}P(r)^2$ where $P(r)$ is proportion of a race population $r$ and $G$ is represented race groups (in our case: white, black, Native American, Asian, Hispanic, Pacific Islander, two or more races, and other races by following ethnicity distribution in the 2010 Census). A higher index number denotes more diversity. However, there is confusion among the general population about the designation of the Hispanic identity since `Hispanic' in the census refers to any `race,' both black and white. So, we decided to use the simple metric, 1 -- white ratio, as `Racial Diversity' since it is clearer. The Pearson correlation between the white ratio and ethnic diversity was 0.667 ($p<0.001$).}, and urbanness: these are our socioeconomic proxy measures. This process resulted in 1,306 users for whom we have self-declared gender and age, as well as inferred income, education level, and characteristics of the area of residence\footnote{We ignored 97 users due to various ZIP code issues, such as ZIP code that were invalid, not available from the census data, or too small to have socioeconomic statistics.}.

In addition to location-derived socioeconomic data, we used news interest as a proxy for socioeconomic variables. According to Pew Research~\cite{pew2012news}, regular news audiences often are more formally educated and have higher household incomes. In particular, readers of The New Yorker and The Economist news media tend to be highly educated and high earners~\cite{pew2012news}. We therefore created a variable that indicates whether each of our users follows The New Yorker ({@}NewYorker) or The Economist ({@}TheEconomist) on Twitter.

\subsection{Genre and Subgenre Information Collection}

\noindent For each user, we extracted the categories of music they listen to at both genre and subgenre (`style') levels. For each user we retrieved the top 50 artists the user listened to via the Last.fm API. We collected genre and subgenre information for each artist using the API for  \textit{Allmusic}\footnote{\url{http://www.allmusic.com/}}, a well-known music database (DB). Unlike other music content databases, Allmusic's metadata is professionally edited and thus is likely to be more consistent when assigning genres or subgenres to artists. Many high-profile music sources like iTunes and Spotify currently use Allmusic to handle relevant artist information.

We matched each artist name collected from Last.fm to an artist entry on the Allmusic DB only if the result exactly matched the queried artist name. When multiple musicians with the same name were matched, we used the Allmusic engine's relevance ranking which is based on usage data and editorial weighting. We manually validated the Allmusic ranking for a random selection of 100 artists that had multiple entries. We examined the Last.fm page for the artist (as linked from the user's Top 50 list, i.e. uniquely identified) and the Allmusic page for the top-ranked artist by the same name as retrieved by the API. We found that the top-ranked artist matches with the Last.fm artist for all cases in this sample of 100.  

A single artist could be classified into multiple genres and subgenres, in which case we distributed the artist's `weight' equally between the respective genres or subgenres. During this data processing, we dropped 292 users who did not have full set of 50 artists that were classified by Allmusic and listened to more than 100 times by the user. As a result, data for 1,014 users were analyzed. There were 8,490 unique artists among the Top 50 artists of 1,014 users, and 987 artists among the unique artists were matched with more than one exact name in Allmusic DB (e.g., Nirvana and Spoon).

\subsection{Measuring Diversity}

\noindent We calculated the diversity of music consumption for each user using both genre- and subgenre-level data derived from their Last.fm activity. We previously argued that in order to explore diversity, we need to investigate multiple factors, namely: the number of genres listened to (variety), the distribution of playing frequency among genres (balance), and, crucially, how related these different genres are (measured via some distance or similarity). These assumptions align well with the concept of Rao-Stirling diversity~\cite{stirling1998economics,stirling2007general,porter2009science,leydesdorff2011indicators}. 

\begin{figure}[tpb!]
\centering
\includegraphics[scale=0.36]{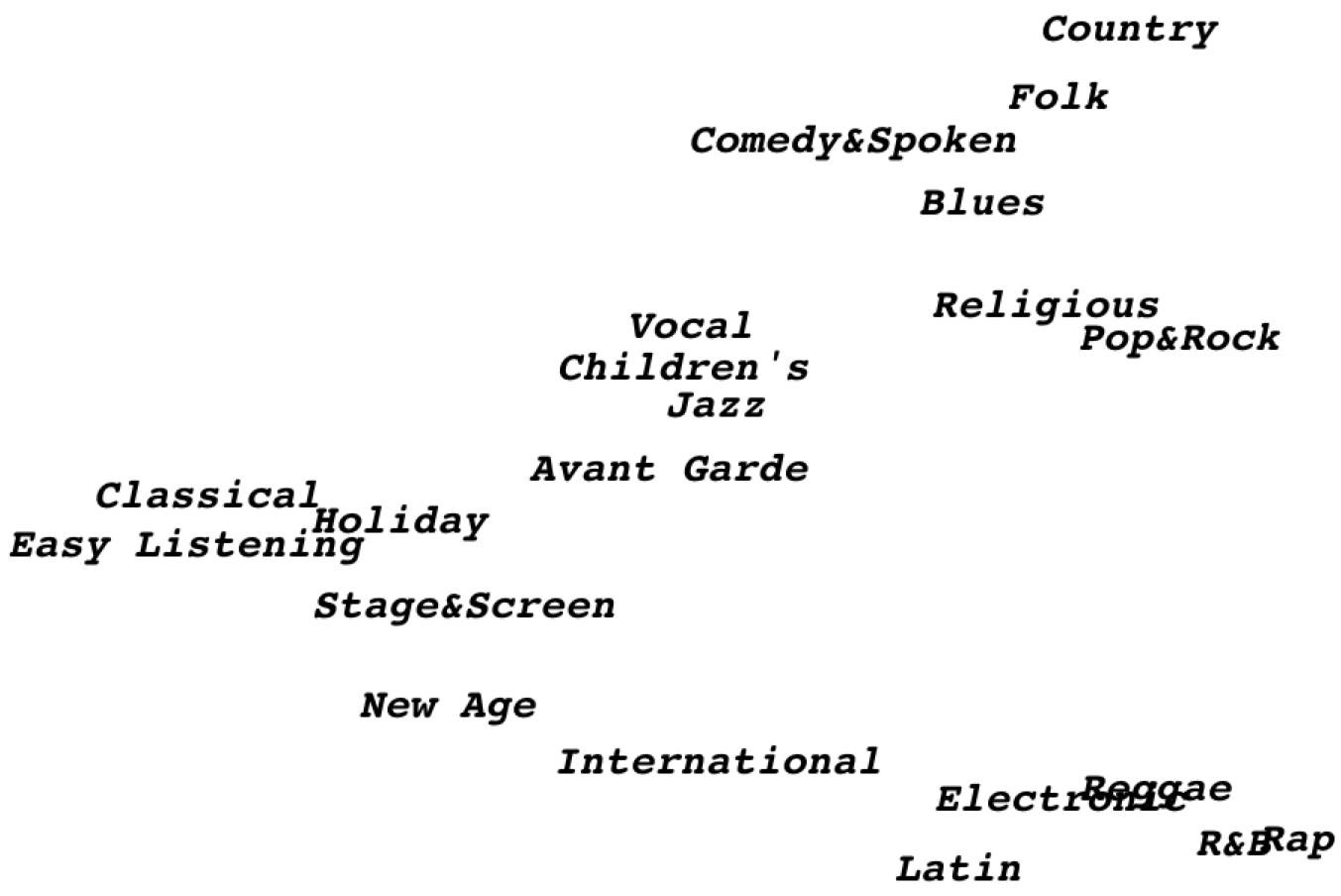}
\caption{Multidimensional scaling for distance between genres}
\label{fig:mds}
\end{figure}

To operationalize the concept of diversity, following Rao-Stirling, we computed the diversity of musical tastes of a user $u$ as $\sum_{i,j \in N} p_{u,i} \times p_{u,j} \times d(i,j)$. In this formulation, $p_{u,i}$ is the fraction of user $u$'s preference for genre $i$ (we performed separate and equivalent calculations for genres and subgenre information; the description here focuses on genre information). To compute $d(i,j)$, we computed the pairwise co-consumption between musical categories as a proxy of closeness. Using an $M \times N$ genre proportion matrix of $p_{u,i}$ values (for each row $u$, $\sum_i p_{u,i} = 1$), we computed every possible pair of genre-to-genre cosine distances between the matrix columns, representing closeness between genres. The distance $d(i,j)$ is the cosine distance, i.e., 1 -- cosine similarity, between the genres. As mentioned above, we repeated the same process with subgenre information. For illustration, the resulting distances for \emph{genres}, embedded in two dimensions using multidimensional scaling~\cite{kruskal1978multidimensional} (MDS), are shown in Figure~\ref{fig:mds}\footnote{Interestingly, highbrow and middlebrow genres (e.g., classical, easy listening, and jazz) are close to each other rather than being close to lowbrow genres (e.g., pop\&rock, folk, country, rap) even though we used an inductive approach to identify the distance between musical categories rather than assuming that musical tastes are shaped by certain schemes.}.

This approach to computing diversity of music consumption has a number of useful qualities. A user who equally (balance) consumes many types of music (variety) that are pairwise highly dissimilar (distance) will have a large diversity score, whereas a user disproportionally consuming a few pairwise similar types of music will has a low diversity score. We evaluate this approach and its robustness below.

\subsection{Into-ness and Openness}

\noindent For each user, we calculated several variables that capture openness (preference for novelty and variety) and into-ness (degree of interest in music) using Twitter and Last.fm data. To help inferring into-ness and openness regarding each user's interests, we first inferred the user's general interests by using a method proposed in \cite{bhattacharya2014inferring}. For a given Twitter user \textit{u} (whose interests are to be inferred), the method first checks which other users \textit{u} is following, i.e., users from whom \textit{u} is interested in receiving information. It then identifies the topics of expertise of those users (whom \textit{u} is following) to infer \textit{u}'s interests, i.e., the topics on which \textit{u} is interested in receiving information. Expertise is defined by the user’s bio or tweets via the Lists feature in Twitter~\cite{ghosh2012cognos}. 

Using the interest topics for each user, we computed openness and into-ness measures. As a proxy of openness, we computed the diversity of the user's interests using the same method we calculated music consumption diversity above. In this case, for example, similarity of interests can be derived from the cosine distance between interest in a matrix that captures users' interest breakdown. As other measures of openness, we counted for each user in our dataset the number of people they are following on Twitter and also the number of unique timezone in 100 randomly sampled people from whom they are following. We collected these openness variables inspired by \cite{schrammel2009personality,quercia2011our}\footnote{We did not consider lexical features of tweets as variables since previous efforts~\cite{golbeck2011predicting,qiu2012you,schwartz2013personality} showed a disagreement regarding predicting features for openness.}.

As a proxy of music into-ness, we used the proportion of music-related interests (any interest topic that included the term `music') among the entire set of user interests along with other types of into-ness that were directly collected via the Last.fm API: number of event attendance in the past, number of loved tracks, period after Last.fm registration, and number of friends in Last.fm.

Table~\ref{tab:hist} presents 15 variables we identified and diversity on genre and subgenre along with their distributions. 

\begin{table}[!t]
\centering
\resizebox{\columnwidth}{!}{%
\begin{tabular}{lll}

\textbf{\fontfamily{pag}\selectfont Socioeconomic Variables} & \textbf{\fontfamily{pag}\selectfont Distribution} & \textbf{\fontfamily{pag}\selectfont Max}\\ \Xhline{2\arrayrulewidth}
{\fontfamily{pag}\selectfont Income} & \raisebox{-.2\height}{\includegraphics[width=3.0cm,height=0.7cm]{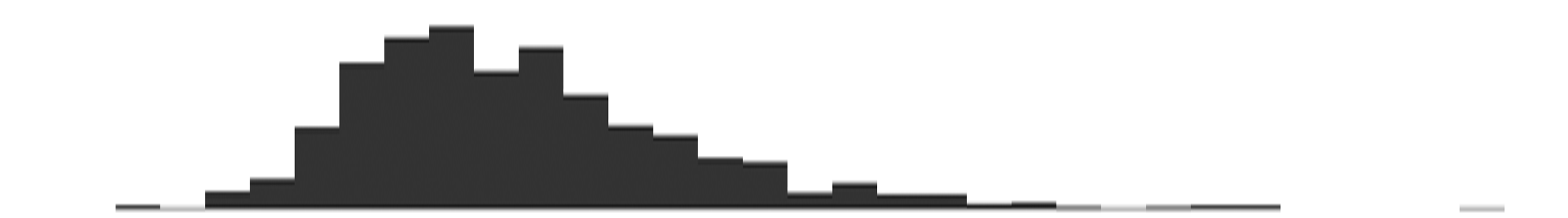}} & {\fontfamily{pag}\selectfont 192,250}\\
{\fontfamily{pag}\selectfont Education} & \raisebox{-.2\height}{\includegraphics[width=3.0cm,height=0.7cm]{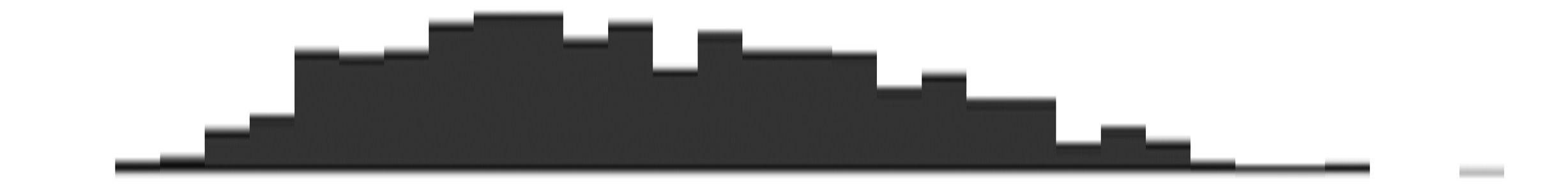}} & {\fontfamily{pag}\selectfont 100}\\ 
{\fontfamily{pag}\selectfont Racial Diversity} & \raisebox{-.2\height}{\includegraphics[width=3.0cm,height=0.7cm]{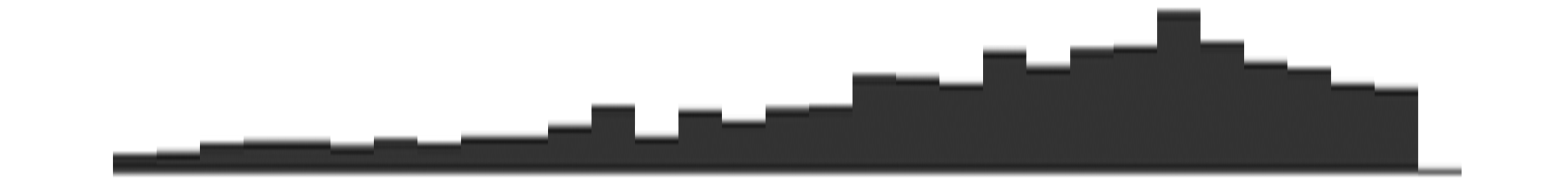}} & {\fontfamily{pag}\selectfont 0.98}\\ 
{\fontfamily{pag}\selectfont High-profile News Reader} & \multicolumn{2}{l}{\raisebox{-.2\height}{\includegraphics[width=5.3cm,height=1cm]{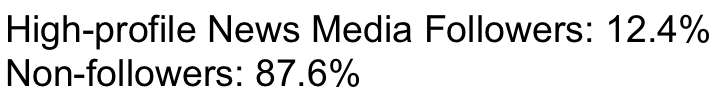}}}\\
{\fontfamily{pag}\selectfont Urbanness} & \raisebox{-.2\height}{\includegraphics[width=3.0cm,height=0.7cm]{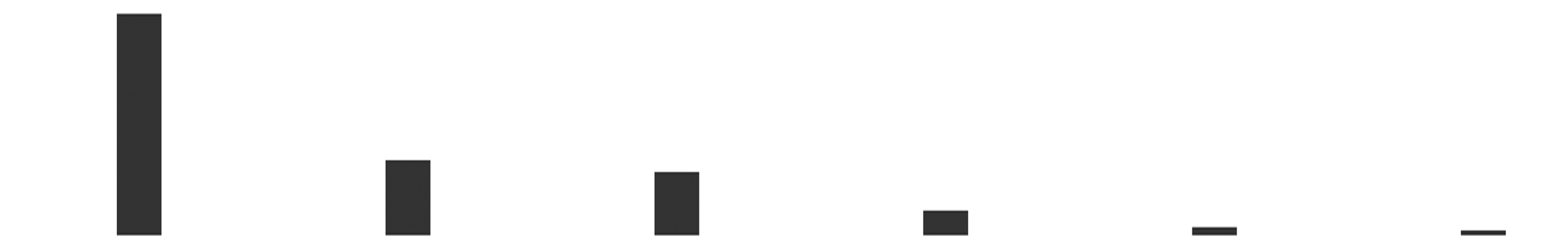}} & {\fontfamily{pag}\selectfont 1--6 (Scale)}\\ \\
\\
\textbf{\fontfamily{pag}\selectfont Demographic Variables} & & \\ \Xhline{2\arrayrulewidth}
{\fontfamily{pag}\selectfont Age} & \raisebox{-.2\height}{\includegraphics[width=3.0cm,height=0.7cm]{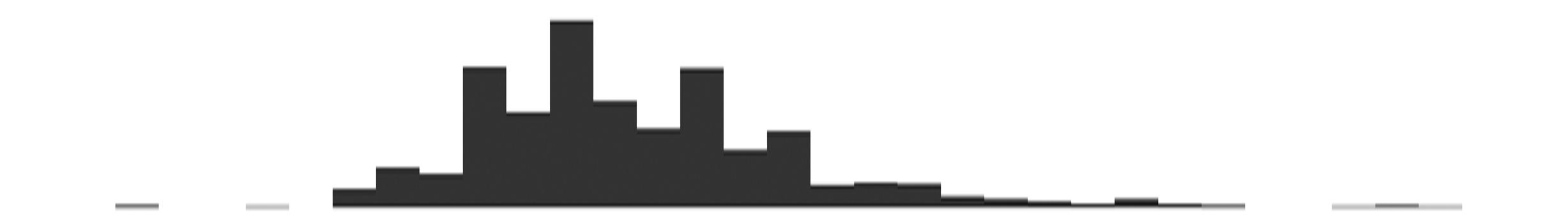}} & {\fontfamily{pag}\selectfont 52}\\
{\fontfamily{pag}\selectfont Gender} & \multicolumn{2}{l}{\raisebox{-.2\height}{\includegraphics[width=5.3cm,height=1cm]{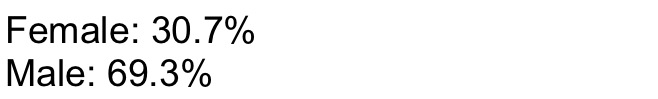}}}\\ \\
\\
\textbf{{\fontfamily{pag}\selectfont Into-ness Variables}} & & \\ \Xhline{2\arrayrulewidth} \\[-1em]
{\fontfamily{pag}\selectfont Musical Event Attendance} & \raisebox{-.2\height}{\includegraphics[width=3.0cm,height=0.7cm]{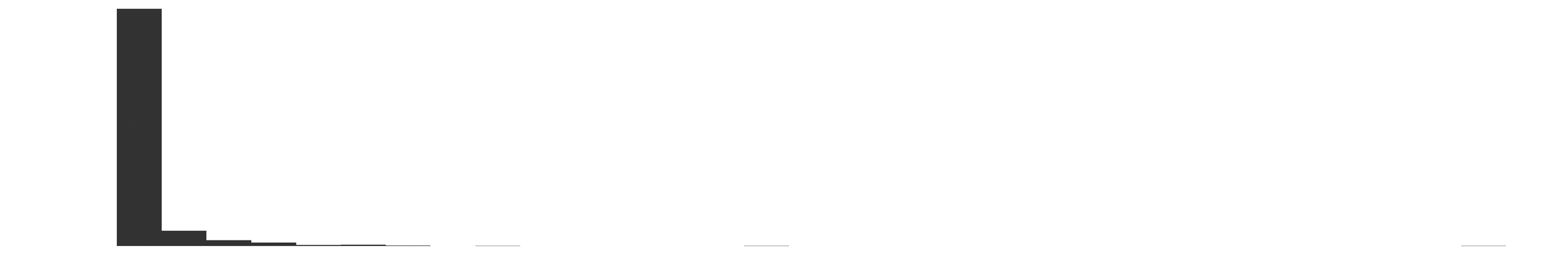}} & {\fontfamily{pag}\selectfont 1,504}\\
{\fontfamily{pag}\selectfont \# of Loved Tracks} & \raisebox{-.2\height}{\includegraphics[width=3.0cm,height=0.7cm]{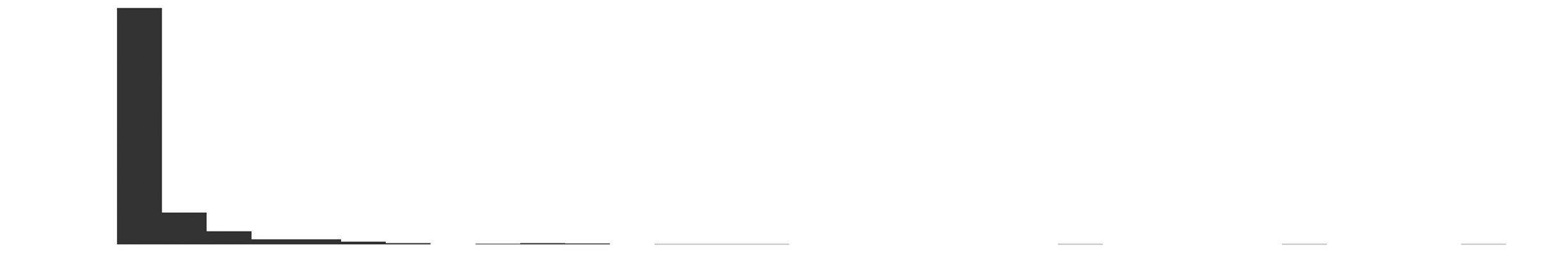}} & {\fontfamily{pag}\selectfont 12,619}\\
{\fontfamily{pag}\selectfont Days from Registration} & \raisebox{-.2\height}{\includegraphics[width=3.0cm,height=0.7cm]{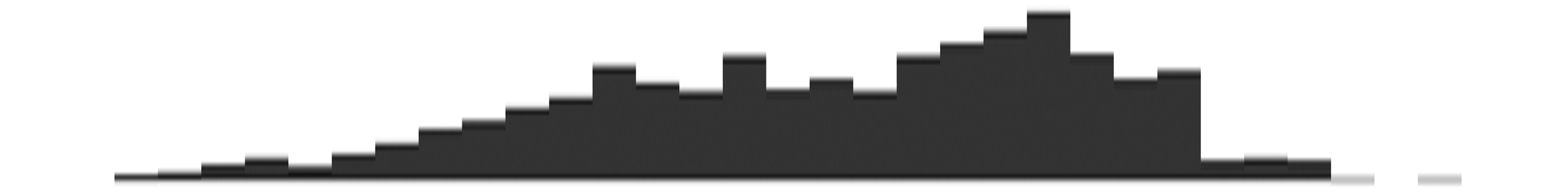}} & {\fontfamily{pag}\selectfont 4,305}\\
{\fontfamily{pag}\selectfont \# of Last.fm Friends} & \raisebox{-.2\height}{\includegraphics[width=3.0cm,height=0.7cm]{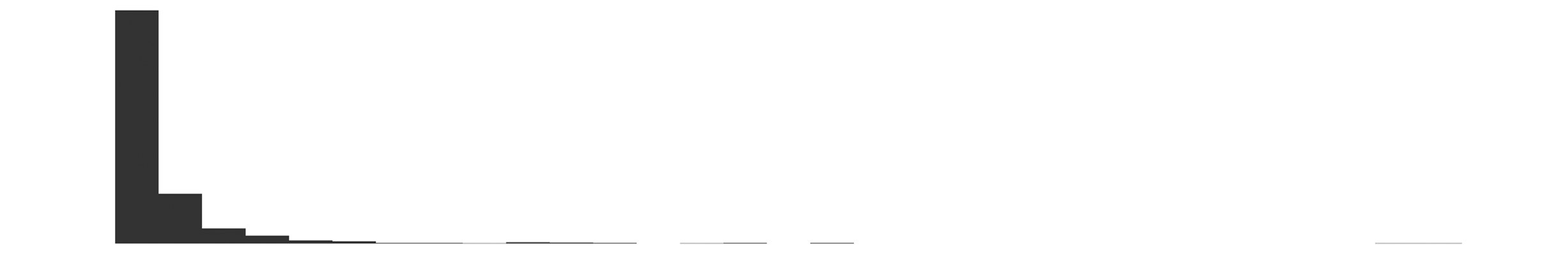}} & {\fontfamily{pag}\selectfont 2,036}\\
{\fontfamily{pag}\selectfont Interest in Music} & \raisebox{-.2\height}{\includegraphics[width=3.0cm,height=0.7cm]{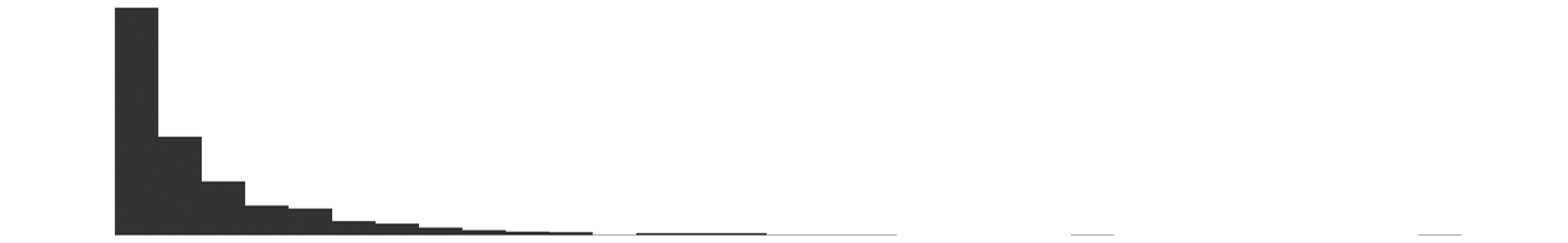}} & {\fontfamily{pag}\selectfont 2,456}\\ \\ 
\\
\textbf{{\fontfamily{pag}\selectfont Openness Variables}} & & \\ \Xhline{2\arrayrulewidth}
{\fontfamily{pag}\selectfont \# of Twitter Friends} & \raisebox{-.2\height}{\includegraphics[width=3.0cm,height=0.7cm]{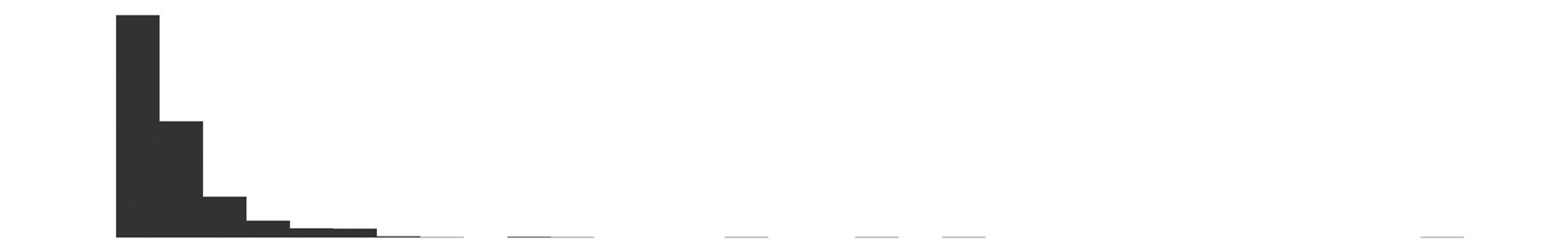}} & {\fontfamily{pag}\selectfont 10,954}\\
{\fontfamily{pag}\selectfont Timezone Diversity of Friends} & \raisebox{-.2\height}{\includegraphics[width=3.0cm,height=0.7cm]{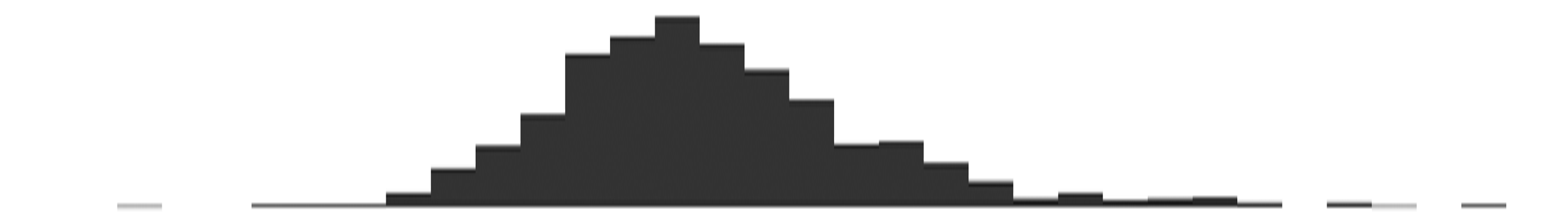}} & {\fontfamily{pag}\selectfont 31}\\
{\fontfamily{pag}\selectfont Interest Diversity} & \raisebox{-.2\height}{\includegraphics[width=3.0cm,height=0.7cm]{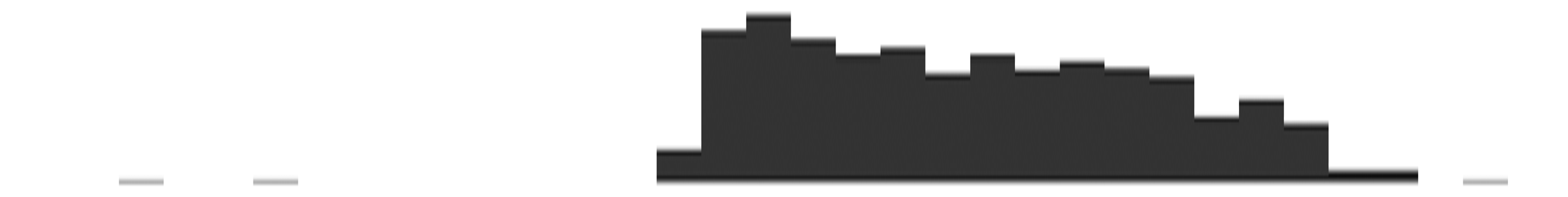}} & {\fontfamily{pag}\selectfont 0.76}\\ \\
\\
\textbf{\fontfamily{pag}\selectfont Diversity} & & \\ \Xhline{2\arrayrulewidth}
{\fontfamily{pag}\selectfont Diversity on Genre} & \raisebox{-.2\height}{\includegraphics[width=3.0cm,height=0.7cm]{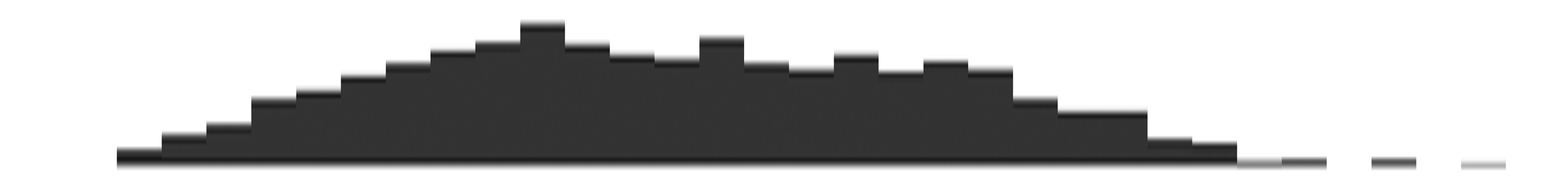}} & {\fontfamily{pag}\selectfont 0.67}\\
{\fontfamily{pag}\selectfont Diversity on Subgenre} & \raisebox{-.2\height}{\includegraphics[width=3.0cm,height=0.7cm]{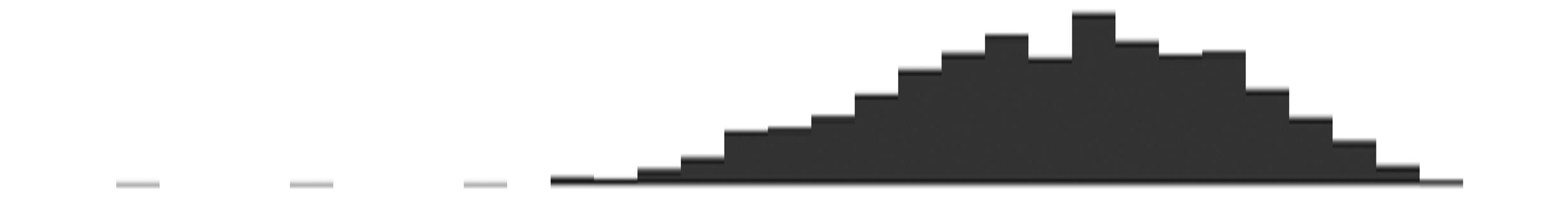}} & {\fontfamily{pag}\selectfont 0.80}\\
\\
\end{tabular}}

\caption{Fifteen variables used to explain the measured musical diversity scores and genre- and subgenre-level of diversity scores. The distributions accompanying each variable begin at zero and end at the adjacent maximum. Many variables are not normally distributed.}
\label{tab:hist}
\end{table}

\subsection{Data Validation and Preparation}

\noindent Given that some of our variables were indirectly derived from social media data, we performed validation tests for our key variables.

\subsubsection{Reverse Geocoding}
To validate our geocoding framework, we matched the inferred ZIP code to the self-reported home location of the user on their Twitter profile. Out of 100 randomly sampled users, eight users did not disclose their location on their Twitter profile or did not properly disclose their location like ``not in a cornfield but\ldots close'' and ``up in the air.'' Among the rest of them (92\% of users), only eight users' locations did not overlap with the inferred zip code location. In other words, more than 90\% of inferred locations were well-matched to the self-reported home locations at town/city/state levels.

Note that it is unusual to have as much as 92\% of users with a valid location field~\cite{hecht2011tweets}. Our dataset, though, includes Twitter users who are also heavy users of the geo-tagged tweets feature; it is conceivable that the same group more readily exposes location in their profile data.

\subsubsection{Socioeconomic Status}
Even if we get the user's location right, the derivation of their socioeconomic information may be wrong as the user may not be \emph{representative} of where they live. For example, it is possible that people who use both Twitter and Last.fm have similar socioeconomic status, regardless of what sort of neighborhood they live in. However, if the inferred socioeconomic information are correct, they should correlate with our other proxy for socioeconomic status: following the New Yorker or Economist. We thus validate our socioeconomic measures by examining whether our inferred income and education level are associated with following the New Yorker ({@}NewYorker) or The Economist ({@}TheEconomist) Twitter accounts. Indeed, compared to other users, New Yorker and Economist followers had higher status for all inferred income and education values, including adjusted gross income (AGI), household income, and level of post-secondary degree (both bachelor's and graduate). These  differences were statistically significant as determined by a one-way ANOVA (New Yorker followers AGI: $p<0.01$; median household income: $p<0.05$; bachelor degree: $p<0.001$; graduate degree: $p<0.001$;  Economist followers AGI: $p<0.001$; median household income: $p<0.05$; bachelor degree: $p<0.001$; graduate degree: $p<0.001$).

\subsubsection{Data Imputation and Standardization}
In our final dataset, 189 out of 1,014 subjects had missing values in one or more variables. According to \cite{hair2006multivariate}, if the missing data level is under 10\% in each variable, any imputation method can be used to augment the missing values. We used multiple imputation methods in our dataset: we applied Bayesian linear regression for continuous variables, and linear discriminant analysis for factor variables. We also standardized all the variables for the final analysis.


\section{Results}

\noindent Our primary purposes for this study were (i) to design a measure that reasonably captures the notion of `diversity of musical tastes' and (ii) to explore associations between musical diversity and various individual factors regarding dimensions of socioeconomic status, demographics, and personal traits including openness and into-ness in music.

\subsection{Diversity Measure}

\noindent To answer RQ1, we estimated the reliability of our diversity measure. We asked three independent annotators to assign a diversity level to the musical consumption of 25 randomly chosen users. The annotators ranged in their music knowledge; we had an expert (musicologist), a music fan, and a causal listener. We provided the annotators two sets of tables of genre- and subgenre-based listening proportion of the 25 users. We asked the annotators to carefully examine each user's listening pattern and apply a 6-point diversity Likert scale where `5' meant very diverse musical taste, `1' meant very low diversity, and `0' meant no diversity at all (it is possible that a user listened only to one genre). We did not provide the annotators with any other information or instructions (such as ``consider the relationship between genres'') as we wanted to know their natural impressions and interpretations of diversity based on their own experiences. Fleiss's Kappa and average pairwise Cohen's Kappa were used to assess the inter-rater reliability for the evaluation. For genre-level the Fleiss Kappa score was 0.411 ($p<0.001$) indicating moderate agreement, and the Cohen's Kappa score was 0.819 ($p<0.001$) indicating almost perfect agreement. For subgenre-level, the respective scores were 0.011 ($p>0.1$) indicating slight agreement and 0.415 ($p<0.05$) indicating moderate agreement. We averaged the rater responses for each user and used that below as the raters' diversity score.

To evaluate our diversity measure, we calculated the Pearson correlation between the raters' average score and our computed diversity score. For genre-level diversity, the correlation between our measure and the raters' diversity was 0.94 ($p<0.001$). For the subgenre-level diversity, the average correlation was 0.87 ($p<0.05$). Interestingly, looking at correlations between individual raters' and our diversity score, the expert annotator had the highest correlation with our diversity score in both settings.  

Other commonly used diversity measures were more sensitive to the level of analysis. We correlated the raters diversity scores with the diversity scores computed by Shannon entropy and by the count of musical categories a user listened to (`volume'). In the genre-level analysis, both the entropy and volume methods showed significant correlation with the raters. The Pearson correlation between the raters' average scores and the entropy values was 0.95 ($p<0.001$). The average correlation between raters and the volume measure was 0.86 ($p<0.001$). However, in subgenre-level analysis we found more notable differences between the raters' and our diversity scores. The Pearson correlations between the entropy and the rater scores was 0.79 ($p<0.05$). With volume, the average correlation was 0.46 ($p<0.05$). 

This result initially indicates that our diversity measure is promising as it captures human rater evaluations of diversity more robustly than traditional measures---it is less dependent on changes in categorical hierarchies. The distance between musical categories can be an important factor for understanding musical diversity, especially in highly complex musical classifications.

\begin{table}[!t] \centering
\resizebox{\columnwidth}{!}{%
\fontfamily{pag}\selectfont
\begin{tabular}{@{\extracolsep{20pt}}lD{.}{.}{-3} D{.}{.}{-3}} 
\\[-1.5ex]\hline 
\hline \\[-1.5ex] 
 & \multicolumn{2}{c}{\textbf{\textit{\fontfamily{pag}\selectfont Dependent variable:}}} \\ 
\cline{2-3} 
\\[-1.5ex] & \multicolumn{1}{c}{\fontfamily{pag}\selectfont \textbf{Genre}} & \multicolumn{1}{c}{\fontfamily{pag}\selectfont \textbf{Subgenre}}\\
\\[-1.5ex] & \multicolumn{1}{c}{\fontfamily{pag}\selectfont \textbf{(1)}} & \multicolumn{1}{c}{\fontfamily{pag}\selectfont \textbf{(2)}}\\ 
\hline \\[-1ex] 
 {\fontfamily{pag}\selectfont Income} & \multicolumn{1}{c}{\fontfamily{pag}\selectfont -0.047} & \multicolumn{1}{c}{\fontfamily{pag}\selectfont 0.007} \\
  & \multicolumn{1}{c}{\fontfamily{pag}\selectfont (0.037)} & \multicolumn{1}{c}{\fontfamily{pag}\selectfont (0.037)} \\ 
  & & \\ 
  {\fontfamily{pag}\selectfont Education} & \multicolumn{1}{c}{\fontfamily{pag}\selectfont 0.027} & \multicolumn{1}{c}{\fontfamily{pag}\selectfont -0.020} \\ 
  & \multicolumn{1}{c}{\fontfamily{pag}\selectfont (0.039)} & \multicolumn{1}{c}{\fontfamily{pag}\selectfont (0.039)} \\ 
  & & \\ 
  {\fontfamily{pag}\selectfont Racial Diversity} & \multicolumn{1}{c}{\fontfamily{pag}\selectfont 0.108$^{**}$} & \multicolumn{1}{c}{\fontfamily{pag}\selectfont 0.089$^{*}$} \\ 
  & \multicolumn{1}{c}{\fontfamily{pag}\selectfont (0.036)} & \multicolumn{1}{c}{\fontfamily{pag}\selectfont (0.036)} \\ 
  & & \\ 
  {\fontfamily{pag}\selectfont Urbanness} & \multicolumn{1}{c}{\fontfamily{pag}\selectfont -0.040} & \multicolumn{1}{c}{\fontfamily{pag}\selectfont 0.052} \\ 
  & \multicolumn{1}{c}{\fontfamily{pag}\selectfont (0.035)} & \multicolumn{1}{c}{\fontfamily{pag}\selectfont (0.035)} \\ 
  & & \\ 
  {\fontfamily{pag}\selectfont High-profile News Reader} & \multicolumn{1}{c}{\fontfamily{pag}\selectfont 0.366$^{***}$} & \multicolumn{1}{c}{\fontfamily{pag}\selectfont 0.301$^{**}$} \\ 
  & \multicolumn{1}{c}{\fontfamily{pag}\selectfont (0.095)} & \multicolumn{1}{c}{\fontfamily{pag}\selectfont (0.096)} \\ 
  & & \\ 
  {\fontfamily{pag}\selectfont Age} & \multicolumn{1}{c}{\fontfamily{pag}\selectfont 0.121$^{***}$} & \multicolumn{1}{c}{\fontfamily{pag}\selectfont 0.161$^{***}$} \\ 
  & \multicolumn{1}{c}{\fontfamily{pag}\selectfont (0.033)} & \multicolumn{1}{c}{\fontfamily{pag}\selectfont (0.033)} \\ 
  & & \\ 
  {\fontfamily{pag}\selectfont Gender (Male)} & \multicolumn{1}{c}{\fontfamily{pag}\selectfont 0.111$^{{\boldsymbol{\cdot}}}$} & \multicolumn{1}{c}{\fontfamily{pag}\selectfont 0.153$^{*}$} \\ 
  & \multicolumn{1}{c}{\fontfamily{pag}\selectfont (0.067)} & \multicolumn{1}{c}{\fontfamily{pag}\selectfont (0.067)} \\ 
  & & \\ 
  {\fontfamily{pag}\selectfont Music Event Attendance} & \multicolumn{1}{c}{\fontfamily{pag}\selectfont -0.145$^{***}$} & \multicolumn{1}{c}{\fontfamily{pag}\selectfont -0.042} \\ 
  & \multicolumn{1}{c}{\fontfamily{pag}\selectfont (0.034)} & \multicolumn{1}{c}{\fontfamily{pag}\selectfont (0.034)} \\ 
  & & \\ 
  {\fontfamily{pag}\selectfont \# of Loved Tracks} & \multicolumn{1}{c}{\fontfamily{pag}\selectfont 0.079$^{*}$} & \multicolumn{1}{c}{\fontfamily{pag}\selectfont 0.089$^{**}$} \\ 
  & \multicolumn{1}{c}{\fontfamily{pag}\selectfont (0.033)} & \multicolumn{1}{c}{\fontfamily{pag}\selectfont (0.033)} \\ 
  & & \\ 
  {\fontfamily{pag}\selectfont Days from Registration} & \multicolumn{1}{c}{\fontfamily{pag}\selectfont -0.102$^{**}$} & \multicolumn{1}{c}{\fontfamily{pag}\selectfont -0.029} \\ 
  & \multicolumn{1}{c}{\fontfamily{pag}\selectfont (0.033)} & \multicolumn{1}{c}{\fontfamily{pag}\selectfont (0.033)} \\ 
  & & \\ 
  {\fontfamily{pag}\selectfont \# of Last.fm Friends} & \multicolumn{1}{c}{\fontfamily{pag}\selectfont 0.023} & \multicolumn{1}{c}{\fontfamily{pag}\selectfont -0.081$^{*}$} \\ 
  & \multicolumn{1}{c}{\fontfamily{pag}\selectfont (0.036)} & \multicolumn{1}{c}{\fontfamily{pag}\selectfont (0.036)} \\ 
  & & \\ 
  {\fontfamily{pag}\selectfont Interest in Music} & \multicolumn{1}{c}{\fontfamily{pag}\selectfont -0.143$^{***}$} & \multicolumn{1}{c}{\fontfamily{pag}\selectfont -0.113$^{***}$} \\ 
  & \multicolumn{1}{c}{\fontfamily{pag}\selectfont (0.034)} & \multicolumn{1}{c}{\fontfamily{pag}\selectfont (0.034)} \\ 
  & & \\ 
  {\fontfamily{pag}\selectfont \# of Twitter Friends} & \multicolumn{1}{c}{\fontfamily{pag}\selectfont 0.085$^{*}$} & \multicolumn{1}{c}{\fontfamily{pag}\selectfont 0.050} \\ 
  & \multicolumn{1}{c}{\fontfamily{pag}\selectfont (0.033)} & \multicolumn{1}{c}{\fontfamily{pag}\selectfont (0.034)} \\ 
  & & \\ 
  {\fontfamily{pag}\selectfont Friends' Timezone Diversity} & \multicolumn{1}{c}{\fontfamily{pag}\selectfont 0.026} & \multicolumn{1}{c}{\fontfamily{pag}\selectfont 0.074$^{*}$} \\ 
  & \multicolumn{1}{c}{\fontfamily{pag}\selectfont (0.032)} & \multicolumn{1}{c}{\fontfamily{pag}\selectfont (0.032)} \\ 
  & & \\ 
  {\fontfamily{pag}\selectfont Interest Diversity} & \multicolumn{1}{c}{\fontfamily{pag}\selectfont -0.027} & \multicolumn{1}{c}{\fontfamily{pag}\selectfont -0.017} \\ 
  & \multicolumn{1}{c}{\fontfamily{pag}\selectfont (0.032)} & \multicolumn{1}{c}{\fontfamily{pag}\selectfont (0.031)} \\ 
  & & \\ 
  {\fontfamily{pag}\selectfont Constant} & \multicolumn{1}{c}{\fontfamily{pag}\selectfont -0.123$^{*}$} & \multicolumn{1}{c}{\fontfamily{pag}\selectfont -0.143$^{*}$} \\ 
  & \multicolumn{1}{c}{\fontfamily{pag}\selectfont (0.057)} & \multicolumn{1}{c}{\fontfamily{pag}\selectfont (0.057)} \\ 
  & & \\ 
\hline \\[-1.5ex] 
Observations & \multicolumn{1}{c}{1,014} & \multicolumn{1}{c}{1,014} \\ 
R$^{2}$ & \multicolumn{1}{c}{0.101} & \multicolumn{1}{c}{0.087} \\ 
Adjusted R$^{2}$ & \multicolumn{1}{c}{0.088} & \multicolumn{1}{c}{0.073} \\ 
Residual Std. Error (df = 998) & \multicolumn{1}{c}{0.955} & \multicolumn{1}{c}{0.963} \\ 
F Statistic (df = 15; 998) & \multicolumn{1}{c}{7.487$^{***}$} & \multicolumn{1}{c}{6.322$^{***}$} \\ 
\hline 
\hline \\[-1.5ex] 
\textit{Note:} & \multicolumn{2}{r}{$^{\boldsymbol{\cdot}}~$p$<$0.1; $^{*}$p$<$0.05; $^{**}$p$<$0.01; $^{*{*}*}$p$<$0.001}\\ 
\end{tabular}}
\caption{Multiple regression coefficients of individual factors on the musical diversity of genre and subgenre}
\label{tab:reg}
\end{table}

\subsection{Correlates of Musical Diversity}

\noindent To address RQ2, we used multiple regression analyses to examine factors associated with the diversity of musical consumption. We examined socioeconomic status variables as well as demographics, openness, and into-ness measures. 

Table~\ref{tab:reg} presents the standardized coefficients of the explanatory variables\footnote{All variance inflation factors are below 1.64 ($\mu=1.28$ and $\sigma=0.16$); Pearson correlation between genre and subgenre diversities is 0.68 ($p<0.001$).}. The model (1) in Table~\ref{tab:reg} estimates the effects of socioeconomic, demographic, and other individual variables on the diversity of musical consumptions on genres. Among the `socioeconomic status' variables, \textit{High-profile News Reader} variable had a high coefficient due to users who follow The Economist or The New Yorker having higher musical diversity than those who do not (one-way ANOVA confirmed the significance; $p<0.001$). Even though we exclude this variable to check whether income and education variables are associated with diversity of music consumption, we could not find any change regarding significance level and direction of correlation. The readers of high-profile news reports may have indirect or subtle difference in terms of socioeconomic status.

\textit{Racial Diversity} positively associates with the diversity of music consumption. This may imply that people in our sample who live in more ethnically diverse area are more likely to have higher musical diversity. By considering the relationship between white ratio and ethnic diversity, this result might be related to the effect of residential segregation. Both of \textit{Age} and \textit{Gender} in the `demographic' variables have positive effect on diversity: being older or male is more likely to have more diverse musical tastes.

Among variables about `into-ness,' \textit{Musical Event Attendance} and \textit{Days from Registration} appear to be negatively associated with diversity, whereas \textit{Number of Last.fm Friends} does not show a significant relationship and \textit{Number of Loved Tracks} appears to positively associated with diversity. \textit{Number of Twitter Friends} as a `openness' variable appears to be positively associated with diversity while \textit{Timezone Diversity of Friends} and \textit{Interest Diversity} shows no effect. On this basis, one could speculate that few variables within the same set of variables correlate with musical diversity in different directions. We discuss these trends below.

Model (2) in Table~\ref{tab:reg} estimates the effects of the same variables on the diversity of musical consumption of subgenres; it shows very similar trends with model (1). However, \textit{Gender} is more significantly associated with diversity. Among the `into-ness' variables, \textit{Number of Last.fm Friends} is significantly associated with diversity rather than \textit{Days from Registration}. But, the general trends of `into-ness' are in common. Among the `openness' variables \textit{Timezone Diversity of Friends} is significantly associated with diversity rather than \textit{Number of Twitter Friends} while the general trends of the `openness' are in common.


\section{Discussion}

\noindent Our results provide initial evidence for the value of our `music diversity measure' which aims to balance three qualities: variety, balance, and distance. Our diversity measure has shown to be more robust than other conventional measures such as volume and entropy. 

Differences between Pearson correlation coefficients at the genre- and subgenre-levels computed by our measure, as well as the average rates assigned by independent coders on a 6-point Likert scale, were not significantly different. For the other measures of diversity, when moving between genre and subgenre levels, the average correlation coefficients dropped more steeply, especially the volume measure. Musical diversity can be computed by simple methods, but it may underestimate or overestimate diversity depending on the complexity of musical categories and the disparity between musical categories that people perceive. Our results show that volume and entropy might not be the best solution for computing the musical diversity of people on a highly complex map of musical categories such as subgenres. 

We only considered the genre and subgenre categories, but new methods for music classification may result in categories that are even more complex, making a robust diversity measure even more important. For example, research efforts have developed novel methods for music classification using various data sources such as audio features and song metadata~\cite{henaff2011unsupervised,foucard2013exploring}. 

In addition, diversity of music consumption was correlated with interest in high-profile news media; users who follow high-profile news media are much more likely to have a higher level of musical diversity. When we think about whether one consumes high-profile news media, it is not necessarily a variable that is as straightforward as income or education level. To understand news reports, readers need more than a basic grasp of word order and word meaning; a particular `knowledge of the world' is also necessary. Van Dijk~\cite{van1996power} explains this when he writes: ``Readers of a news report first of all need to understand its words, sentences, or the structural properties. This does not only mean they must know the language and its grammar and lexicon, possibly including rather technical words such as those of modern politics, management, science, or the professions. Users of the media need to know something about the specific organization and functions of news reports in the press, including the functions of headlines, leads, background information, or quotations. Besides such grammatical and textual knowledge, media users need vast amounts of properly organized knowledge of the world.'' Van Dijk's point alludes to the possibility that if one has access to particular understandings of `the world,' then they are better equipped to seek out and benefit from high profile news sources. If this is the case, then we can begin to think about level of music diversity as a potential variable vis-\`a-vis knowledge.

Our results also confirm a number of previous findings about demographic variables associated with the diversity of music consumption. They show that male users are more likely to have diverse musical tastes, which confirms prior research showing that males tend to consider mainstream music as \textit{unhip} while females consider it in another way of saying \textit{popular} music~\cite{christenson1988genre}; such perceptions might affect musical consumption. Males are also more likely to prefer more unique styles of music than females~\cite{rawlings1997music}. In addition, people who are older in our sample are more likely to have diverse musical tastes. This result closely echos the analyses of \cite{warde2007understanding}: young people may identify strongly with one or only a few genres and styles of music, which reveals the significance of their representational dimensions.

A potentially surprising finding is that people who attended more musical events are less likely to have diverse listening habits. \cite{warde2007understanding} also argues that there is a tendency, or an openness, towards unfamiliar musical forms and evidence of relatively diverse tastes in people who are limited in how they can engage in musical activities. The development of a broad palette of musical tastes was not valued by people for whom music is more accessible. We note that urban dwellers may have better access to musical activities, but we could not find a significant association with urbanness in our results.

Users with diverse patterns of music consumption are less likely to follow music-related accounts on Twitter. This finding can be due to a different set of music-related interests between diverse listeners---who care more about the music itself---and more casual music fans who may care more about the celebrity factor. If this were shown to be true, we may refer to it as the Justin Bieber effect (no offense to his fans should they be reading this paper).

\section{Final Remarks}

\noindent In this paper, we have designed a reasonable measure that quantifies the diversity of musical tastes. In addition, we provide an analysis of diversity as it relates to the cultural omnivore thesis. Based on well-known individual factors which relate diversity of musical preferences across various theoretical work and empirical studies, we identified key factors for designing a diversity measure, and located individual-level variables for exploring correlations of musical diversity. 

We acknowledge that the manner in which we inferred the socioeconomic status variable could produce significant inaccuracies. For example, users' home locations were inferred in ZIP code resolution and using geocoded Twitter data. These methods are prone to error. Other methods for collecting more direct or fine-grained location data, or maybe even a direct collection of socioeconomic variables, might give us a better opportunity to study this correlation with music consumption. Second, our user population and the music they listen to are both potentially highly biased. Our population is comprised of users who make an explicit connection between their Twitter and Last.fm accounts, which may indicate search biases on our behalf. In addition, the tracks and artists displayed for each user are based on their public listening behavior, which may or may not be reflective of their overall listening habits. Finally, we could see rating differences among coders due to knowledge differences. The measurement validations can be improved by better systematic investigations using more listening history samples and annotators with different levels of knowledge background. At the same time, it would be interesting to see if the way of rating changes when the music listeners themselves are asked about their diversity.

Future research along this vein will provide a richer and more complex picture of musical preferences. This picture will in turn contribute to a greater understanding of the changing face of the cultural omnivore, as it manifests through analyses of social media data, and also contribute to a empirical recommendation system aiming to provide contents based on tastes and aesthetics preferences.

\section{Acknowledgements}

\noindent We are grateful to Gabriel Magno, Luam Cat\~ao, Kunwoo Park, Jaehyuk Park, Seungback Shin, Amit Sharma, Nir Grinberg, Michael Macy, and Dan Cosley for their thoughtful questions, comments, and helps on this study, and we also appreciate the anonymous reviewers for their valuable comments.

\bibliographystyle{aaai}
\bibliography{references}

\begin{thebibliography}{}

\bibitem[\protect\citeauthoryear{Bhattacharya \bgroup et al\mbox.\egroup
  }{2014}]{bhattacharya2014inferring}
Bhattacharya, P.; Zafar, M.~B.; Ganguly, N.; Ghosh, S.; and Gummadi, K.~P.
\newblock 2014.
\newblock Inferring user interests in the twitter social network.
\newblock In {\em ACM Conference on Recommender systems (RecSys)},  357--360.

\bibitem[\protect\citeauthoryear{Bourdieu}{1984}]{bourdieu1984distinction}
Bourdieu, P.
\newblock 1984.
\newblock {\em Distinction: A social critique of the judgement of taste}.
\newblock Harvard University Press.

\bibitem[\protect\citeauthoryear{Bu \bgroup et al\mbox.\egroup
  }{2010}]{bu2010music}
Bu, J.; Tan, S.; Chen, C.; Wang, C.; Wu, H.; Zhang, L.; and He, X.
\newblock 2010.
\newblock Music recommendation by unified hypergraph: Combining social media
  information and music content.
\newblock In {\em ACM International Conference on Multimedia (ACMMM)},
  391--400.

\bibitem[\protect\citeauthoryear{Buld{\'u} \bgroup et al\mbox.\egroup
  }{2007}]{buldu2007complex}
Buld{\'u}, J.~M.; Cano, P.; Koppenberger, M.; Almendral, J.~A.; and Boccaletti,
  S.
\newblock 2007.
\newblock The complex network of musical tastes.
\newblock {\em New Journal of Physics} 9(6):172.

\bibitem[\protect\citeauthoryear{Castelli \bgroup et al\mbox.\egroup
  }{2009}]{castelli2009extracting}
Castelli, G.; Mamei, M.; Rosi, A.; and Zambonelli, F.
\newblock 2009.
\newblock Extracting high-level information from location data: The w4 diary
  example.
\newblock {\em Mobile Networks and Applications} 14(1):107--119.

\bibitem[\protect\citeauthoryear{Chen, Wu, and He}{2013}]{chen2013personality}
Chen, L.; Wu, W.; and He, L.
\newblock 2013.
\newblock How personality influences users' needs for recommendation diversity?
\newblock In {\em CHI'13 Extended Abstracts on Human Factors in Computing
  Systems},  829--834.

\bibitem[\protect\citeauthoryear{Christenson and
  Peterson}{1988}]{christenson1988genre}
Christenson, P.~G., and Peterson, J.~B.
\newblock 1988.
\newblock Genre and gender in the structure of music preferences.
\newblock {\em Communication Research} 15(3):282--301.

\bibitem[\protect\citeauthoryear{Coulangeon and
  Lemel}{2007}]{coulangeon2007distinction}
Coulangeon, P., and Lemel, Y.
\newblock 2007.
\newblock Is `distinction' really outdated? questioning the meaning of the
  omnivorization of musical taste in contemporary france.
\newblock {\em Poetics} 35(2):93--111.

\bibitem[\protect\citeauthoryear{DeNora}{2000}]{denora2000music}
DeNora, T.
\newblock 2000.
\newblock {\em Music in everyday life}.
\newblock Cambridge University Press.

\bibitem[\protect\citeauthoryear{Farrahi \bgroup et al\mbox.\egroup
  }{2014}]{farrahi2014impact}
Farrahi, K.; Schedl, M.; Vall, A.; Hauger, D.; and Tkalcic, M.
\newblock 2014.
\newblock Impact of listening behavior on music recommendation.
\newblock In {\em International Society for Music Information Retrieval
  Conference (ISMIR)}.

\bibitem[\protect\citeauthoryear{Foucard \bgroup et al\mbox.\egroup
  }{2013}]{foucard2013exploring}
Foucard, R.; Essid, S.; Richard, G.; and Lagrange, M.
\newblock 2013.
\newblock Exploring new features for music classification.
\newblock In {\em IEEE International Workshop on Image Analysis for Multimedia
  Interactive Services (WIAMIS)},  1--4.

\bibitem[\protect\citeauthoryear{Ghosh \bgroup et al\mbox.\egroup
  }{2012}]{ghosh2012cognos}
Ghosh, S.; Sharma, N.; Benevenuto, F.; Ganguly, N.; and Gummadi, K.
\newblock 2012.
\newblock Cognos: Crowdsourcing search for topic experts in microblogs.
\newblock In {\em International ACM SIGIR conference on Research and
  Development in Information Retrieval},  575--590.

\bibitem[\protect\citeauthoryear{Golbeck \bgroup et al\mbox.\egroup
  }{2011}]{golbeck2011predicting}
Golbeck, J.; Robles, C.; Edmondson, M.; and Turner, K.
\newblock 2011.
\newblock Predicting personality from twitter.
\newblock In {\em IEEE International Conference on Social Computing
  (SocialCom)},  149--156.

\bibitem[\protect\citeauthoryear{Goldberg}{2011}]{goldberg2011mapping}
Goldberg, A.
\newblock 2011.
\newblock Mapping shared understandings using relational class analysis: The
  case of the cultural omnivore reexamined1.
\newblock {\em American Journal of Sociology} 116(5):1397--1436.

\bibitem[\protect\citeauthoryear{Hair \bgroup et al\mbox.\egroup
  }{2006}]{hair2006multivariate}
Hair, J.~F.; Tatham, R.~L.; Anderson, R.~E.; and Black, W.
\newblock 2006.
\newblock {\em Multivariate data analysis}, volume~6.
\newblock Pearson Prentice Hall Upper Saddle River, NJ.

\bibitem[\protect\citeauthoryear{Hecht and Stephens}{2014}]{hecht2014tale}
Hecht, B., and Stephens, M.
\newblock 2014.
\newblock A tale of cities: Urban biases in volunteered geographic information.
\newblock In {\em AAAI conference on Weblogs and Social Media (ICWSM)}.

\bibitem[\protect\citeauthoryear{Hecht \bgroup et al\mbox.\egroup
  }{2011}]{hecht2011tweets}
Hecht, B.; Hong, L.; Suh, B.; and Chi, E.~H.
\newblock 2011.
\newblock Tweets from justin bieber's heart: The dynamics of the location field
  in user profiles.
\newblock In {\em ACM SIGCHI Conference on Human Factors in Computing Systems
  (CHI)},  237--246.

\bibitem[\protect\citeauthoryear{Henaff \bgroup et al\mbox.\egroup
  }{2011}]{henaff2011unsupervised}
Henaff, M.; Jarrett, K.; Kavukcuoglu, K.; and LeCun, Y.
\newblock 2011.
\newblock Unsupervised learning of sparse features for scalable audio
  classification.
\newblock In {\em International Society of Music Information Retrieval
  (ISMIR)},  681--686.

\bibitem[\protect\citeauthoryear{Hurley and Zhang}{2011}]{hurley2011novelty}
Hurley, N., and Zhang, M.
\newblock 2011.
\newblock Novelty and diversity in top-n recommendation--analysis and
  evaluation.
\newblock {\em ACM Transactions on Internet Technology (TOIT)} 10(4):14.

\bibitem[\protect\citeauthoryear{Kruskal and
  Wish}{1978}]{kruskal1978multidimensional}
Kruskal, J.~B., and Wish, M.
\newblock 1978.
\newblock {\em Multidimensional scaling}, volume~11.
\newblock Sage.

\bibitem[\protect\citeauthoryear{Lewis, Gonzalez, and
  Kaufman}{2012}]{lewis2012social}
Lewis, K.; Gonzalez, M.; and Kaufman, J.
\newblock 2012.
\newblock Social selection and peer influence in an online social network.
\newblock {\em Proceedings of the National Academy of Sciences} 109(1):68--72.

\bibitem[\protect\citeauthoryear{Leydesdorff and
  Rafols}{2011}]{leydesdorff2011indicators}
Leydesdorff, L., and Rafols, I.
\newblock 2011.
\newblock Indicators of the interdisciplinarity of journals: Diversity,
  centrality, and citations.
\newblock {\em Journal of Informetrics} 5(1):87--100.

\bibitem[\protect\citeauthoryear{McNee, Riedl, and
  Konstan}{2006}]{mcnee2006being}
McNee, S.~M.; Riedl, J.; and Konstan, J.~A.
\newblock 2006.
\newblock Being accurate is not enough: How accuracy metrics have hurt
  recommender systems.
\newblock In {\em CHI'06 Extended Abstracts on Human Factors in Computing
  Systems},  1097--1101.

\bibitem[\protect\citeauthoryear{M{\"o}rchen \bgroup et al\mbox.\egroup
  }{2005}]{morchen2005databionic}
M{\"o}rchen, F.; Ultsch, A.; N{\"o}cker, M.; and Stamm, C.
\newblock 2005.
\newblock Databionic visualization of music collections according to perceptual
  distance.
\newblock In {\em International Society of Music Information Retrieval
  (ISMIR)},  396--403.

\bibitem[\protect\citeauthoryear{Peterson}{1992}]{peterson1992understanding}
Peterson, R.~A.
\newblock 1992.
\newblock Understanding audience segmentation: From elite and mass to omnivore
  and univore.
\newblock {\em Poetics} 21(4):243--258.

\bibitem[\protect\citeauthoryear{Peterson}{1997}]{peterson1997rise}
Peterson, R.~A.
\newblock 1997.
\newblock The rise and fall of highbrow snobbery as a status marker.
\newblock {\em Poetics} 25(2):75--92.

\bibitem[\protect\citeauthoryear{Peterson}{2005}]{peterson2005problems}
Peterson, R.~A.
\newblock 2005.
\newblock Problems in comparative research: The example of omnivorousness.
\newblock {\em poetics} 33(5):257--282.

\bibitem[\protect\citeauthoryear{{Pew Research Center}}{2012}]{pew2012news}
{Pew Research Center}.
\newblock 2012.
\newblock Demographics and political views of news audiences.
\newblock \url{http://tinyurl.com/kbb6n9m}.
\newblock [Online; accessed 30-December-2014].

\bibitem[\protect\citeauthoryear{Porter and Rafols}{2009}]{porter2009science}
Porter, A.~L., and Rafols, I.
\newblock 2009.
\newblock Is science becoming more interdisciplinary? measuring and mapping six
  research fields over time.
\newblock {\em Scientometrics} 81(3):719--745.

\bibitem[\protect\citeauthoryear{Qiu \bgroup et al\mbox.\egroup
  }{2012}]{qiu2012you}
Qiu, L.; Lin, H.; Ramsay, J.; and Yang, F.
\newblock 2012.
\newblock You are what you tweet: Personality expression and perception on
  twitter.
\newblock {\em Journal of Research in Personality} 46(6):710--718.

\bibitem[\protect\citeauthoryear{Quercia \bgroup et al\mbox.\egroup
  }{2011}]{quercia2011our}
Quercia, D.; Kosinski, M.; Stillwell, D.; and Crowcroft, J.
\newblock 2011.
\newblock Our twitter profiles, our selves: Predicting personality with
  twitter.
\newblock In {\em IEEE International Conference on Social Computing
  (SocialCom)},  180--185.

\bibitem[\protect\citeauthoryear{Rawlings and
  Ciancarelli}{1997}]{rawlings1997music}
Rawlings, D., and Ciancarelli, V.
\newblock 1997.
\newblock Music preference and the five-factor model of the neo personality
  inventory.
\newblock {\em Psychology of Music} 25(2):120--132.

\bibitem[\protect\citeauthoryear{Rentfrow and Gosling}{2003}]{rentfrow2003re}
Rentfrow, P.~J., and Gosling, S.~D.
\newblock 2003.
\newblock The do re mi's of everyday life: The structure and personality
  correlates of music preferences.
\newblock {\em Journal of personality and social psychology} 84(6):1236.

\bibitem[\protect\citeauthoryear{Rimmer}{2012}]{rimmer2012beyond}
Rimmer, M.
\newblock 2012.
\newblock Beyond omnivores and univores: The promise of a concept of musical
  habitus.
\newblock {\em Cultural Sociology} 6(3):299--318.

\bibitem[\protect\citeauthoryear{Schrammel, K{\"o}ffel, and
  Tscheligi}{2009}]{schrammel2009personality}
Schrammel, J.; K{\"o}ffel, C.; and Tscheligi, M.
\newblock 2009.
\newblock Personality traits, usage patterns and information disclosure in
  online communities.
\newblock In {\em British HCI Group Annual Conference on People and Computers:
  Celebrating People and Technology},  169--174.

\bibitem[\protect\citeauthoryear{Schwartz \bgroup et al\mbox.\egroup
  }{2013}]{schwartz2013personality}
Schwartz, H.~A.; Eichstaedt, J.~C.; Kern, M.~L.; Dziurzynski, L.; Ramones,
  S.~M.; Agrawal, M.; Shah, A.; Kosinski, M.; Stillwell, D.; Seligman, M.~E.;
  et~al.
\newblock 2013.
\newblock Personality, gender, and age in the language of social media: The
  open-vocabulary approach.
\newblock {\em PloS One} 8(9):e73791.

\bibitem[\protect\citeauthoryear{Stirling}{1998}]{stirling1998economics}
Stirling, A.
\newblock 1998.
\newblock On the economics and analysis of diversity.
\newblock {\em Science Policy Research Unit (SPRU), Electronic Working Papers
  Series} 28:1--156.

\bibitem[\protect\citeauthoryear{Stirling}{2007}]{stirling2007general}
Stirling, A.
\newblock 2007.
\newblock A general framework for analysing diversity in science, technology
  and society.
\newblock {\em Journal of the Royal Society Interface} 4(15):707--719.

\bibitem[\protect\citeauthoryear{Turnbull \bgroup et al\mbox.\egroup
  }{2014}]{turnbull2014using}
Turnbull, D.~R.; Zupnick, J.~A.; Stensland, K.~B.; Horwitz, A.~R.; Wolf, A.~J.;
  Spirgel, A.~E.; Meyerhofer, S.~P.; and Joachims, T.
\newblock 2014.
\newblock Using personalized radio to enhance local music discovery.
\newblock In {\em CHI'14 Extended Abstracts on Human Factors in Computing
  Systems},  2023--2028.

\bibitem[\protect\citeauthoryear{Van~Dijk}{1996}]{van1996power}
Van~Dijk, T.~A.
\newblock 1996.
\newblock Power and the news media.
\newblock {\em Political communication in action}  9--36.

\bibitem[\protect\citeauthoryear{Wallin, Merker, and
  Brown}{2001}]{wallin2001origins}
Wallin, N.~L.; Merker, B.; and Brown, S.
\newblock 2001.
\newblock {\em The origins of music}.
\newblock MIT press.

\bibitem[\protect\citeauthoryear{Warde, Wright, and
  Gayo-Cal}{2007}]{warde2007understanding}
Warde, A.; Wright, D.; and Gayo-Cal, M.
\newblock 2007.
\newblock Understanding cultural omnivorousness: Or, the myth of the cultural
  omnivore.
\newblock {\em Cultural sociology} 1(2):143--164.

\bibitem[\protect\citeauthoryear{{Wikipedia}}{2015}]{wiki2015lastfm}
{Wikipedia}.
\newblock 2015.
\newblock Last.fm.
\newblock \url{http://en.wikipedia.org/wiki/Last.fm}.
\newblock [Online; accessed 19-January-2015].

\bibitem[\protect\citeauthoryear{Yang, Wang, and
  Mourali}{2014}]{yang2014effect}
Yang, Z.; Wang, J.; and Mourali, M.
\newblock 2014.
\newblock Effect of peer influence on unauthorized music downloading and
  sharing: The moderating role of self-construal.
\newblock {\em Journal of Business Research}.

\bibitem[\protect\citeauthoryear{Zheleva \bgroup et al\mbox.\egroup
  }{2010}]{zheleva2010statistical}
Zheleva, E.; Guiver, J.; Mendes~Rodrigues, E.; and Mili{\'c}-Frayling, N.
\newblock 2010.
\newblock Statistical models of music-listening sessions in social media.
\newblock In {\em International World Wide Web Conference (WWW)},  1019--1028.

\end{thebibliography}
\end{document}